\documentclass[12pt]{article}
\usepackage{epsfig}

 \usepackage{latexsym}
 \usepackage{amssymb}

\usepackage{bm}

\newcommand{\be}{\begin{equation}}
\newcommand{\ee}{\end{equation}}
\newcommand{\bear}{\be\begin{array}}
\newcommand{\bea}{\begin{eqnarray}}
\newcommand{\eea}{\end{eqnarray}}

\newcommand{\dst}{\displaystyle}
\newcommand{\fr}[2]{\frac{{\dst #1}}{{\dst #2}}}

 \date{October 27, 2005}

\title{MUON PAIR PRODUCTION IN RELATIVISTIC NUCLEAR
COLLISIONS\footnote{Report presented by V.G.~Serbo at the
International Conference ``The PHOTON: its first hundred years and
the future'' (30.08 --- 08.09.2005, Warsaw and Kazimierz, Poland).
This work is supported in part by RFBR (code 05-02-16211) and by the
Fund of Russian Scientific Schools (code 2339.2003.2). V.G.S.
acknowledged the financial support from the Organizing Committee.}}

\author{K.~Hencken$^{1)}$, E.A.~Kuraev$^{2)}$, V.G.~Serbo$^{3)}$\\
{\it $^{1)}$University of Basel, 4056, Basel, Switzerland}\\
{\it  $^{2)}$Joint Institute of Nuclear Reseach, 141980, Dubna,
Russia}\\
{\it $^{3)}$Novosibirsk State University, 630090, Novosibirsk,
Russia}}

\begin{document}

\maketitle

\begin{abstract}
The exclusive production of one $\mu^+\mu^-$ pair in collisions of
two ultra-relativistic nuclei is considered. We present the simple
method for calculation of the Born cross section for this process.
Then we found that the Coulomb corrections to this cross section
(which correspond to multi-photon exchange of the produced
$\mu^{\pm}$ with nuclei) are small while the unitarity corrections
are large. This is in sharp contrast to the exclusive $e^+e^-$ pair
production where the Coulomb corrections to the Born cross section
are large while the unitarity corrections are small. We calculated
also the cross section for the production of one $\mu^+\mu^-$ pair
and several $e^+e^-$ pairs in the leading logarithmic approximation.
Using this cross section we found that the inclusive production of
$\mu^+\mu^-$ pair coincides in this approximation with its Born
value.
\end{abstract}

\section{Introduction}

The lepton pair production in ultra-relativistic nuclear collisions
were discussed in numerous papers (see review~\cite{BHTSK} and
references therein). For the RHIC and LHC colliders the charge
numbers of nuclei $Z_1=Z_2\equiv Z$ and their Lorentz factors
$\gamma_1=\gamma_2\equiv \gamma$ are given in Table ~\ref{t1}.

\begin{table}[!h]
\vspace{5mm} {\renewcommand{\arraystretch}{1.5}
  \caption{Colliders and the Born cross sections for lepton pair
production}
\begin{center}
\par
 \begin{tabular}{|c|c|c|c|c|c|}\hline
Collider & $Z$ & $\gamma$ & $\sigma^{e^+e^-}_{\rm Born}$ [kb]
&$\sigma^{\mu^+\mu^-}_{\rm Born}$ [b]
\\ \hline
RHIC, Au-Au & 79 & 108 & 36.0 & 0.23
\\ \hline
LHC, Pb-Pb & 82 & 3000 & 227 & 2.6
\\ \hline
\end{tabular}
 \label{t1}
\end{center}
 }
 \end{table}

\begin{figure}[!h]
\begin{center}
\includegraphics[width=3.5cm,angle=0]{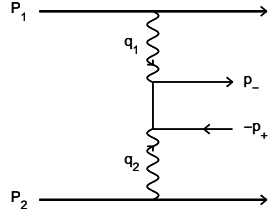}
 \caption{The Feynman diargam for the lepton pair
 production in the Born approximation}
\end{center}
 \label{F:1}
\end{figure}

The cross section of one $e^+e^-$ pair production in the Born
approximation, described by Feynman diagram of Fig. \ref{F:1}, was
obtained many years ago~\cite{Landau}. Since the Born cross section
$\sigma^{e^+e^-}_{\rm Born}$ is huge (see Table~\ref{t1}), the
$e^+e^-$  pair production can be a serious background for many
experiments. It is also important for the problem of beam lifetime
and luminosity of colliders. It means that the various corrections
to the Born cross section are of great importance. At present, there
are a lot of controversial and incorrect statements in papers
devoted to this subject. The corresponding references and critical
remarks can be found in~\cite{BHTSK,ISS-99,LMS-02}.
Since the parameter  $Z\alpha$ may be not small ($Z\alpha \approx
0.6$ for Au-Au and Pb-Pb collisions), the whole series in $Z\alpha$
has to be summed to obtain the cross section with sufficient
accuracy. The exact cross section for one pair production $\sigma_1$
can be written in the form
 \be
\sigma_1 = \sigma_{\rm Born} + \sigma_{\rm Coul}+ \sigma_{\rm
unit}\,,
 \label{1}
 \ee
where two different types of corrections have to be distinguished.
\begin{figure}[!h]
\begin{center}
\includegraphics[width=3.5cm,angle=0]{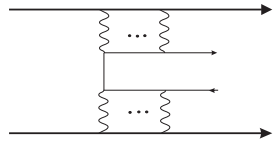}
 \caption{The Feynman diargam for the Coulomb correction}
  \label{F:2}
\end{center}
\end{figure}
The Coulomb correction $\sigma_{\rm Coul}$ corresponds to
multi-photon exchange of the produced $e^{\pm}$ with nuclei (Fig.
\ref{F:2}); it was calculated in~\cite{ISS-99}. The unitarity
correction $\sigma_{\rm unit}$ corresponds to the exchange of
light-by-light blocks between nuclei (Fig. \ref{F:3}); it was
calculated in~\cite{LMS-02}. The results of~\cite{LMS-02} recently
were confirmed in~\cite{BGKN} by direct summation of the Feynman
diagrams.
\begin{figure}[!h]
\begin{center}
\includegraphics[width=3.5cm,angle=0]{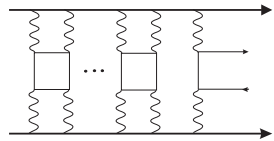}
 \caption{The Feynman diargam for the unitarity correction}
  \label{F:3}
\end{center}
\end{figure}
It was found that the Coulomb corrections are large while the
unitarity corrections are small (see Table~\ref{t2}).
\begin{table}[!h]
\vspace{5mm} {\renewcommand{\arraystretch}{1.5}
 \caption{Coulomb and unitarity corrections to the $e^+e^-$ pair
 production}
\begin{center}
\par
  \begin{tabular}{|c|c|c|c|}\hline
Collider & $\frac{\displaystyle\sigma_{\rm Coul}}{\displaystyle
\sigma_{\rm Born}}$ & $\frac{\displaystyle\sigma_{\rm
unit}}{\displaystyle
\sigma_{\rm Born}}$ 
\\ \hline
RHIC, Au-Au &  $-25$\% & $-4.1$\%
\\ \hline
LHC, Pb-Pb & $-14$\% & $-3.3$\%
\\ \hline
\end{tabular}
 \label{t2}
\end{center}
 }
\end{table}

Muon pair production may be easier for an experimental
observation. The calculation scheme for the $\mu^+\mu^-$ pair
production is quite different from that for the $e^+e^-$ pair
production.

\section{Born cross section for one $\mu^+\mu^-$ pair
production}

The production of one {$\mu^{+}\mu^{-}$} pair
  \be
Z_1+Z_2 \to Z_1+Z_2+ \mu^+\mu^-
 \label{2}
 \ee
in the Born approximation is described by the Feynman diagram of
Fig. 1. When two nuclei with charges $Z_1e$ and $Z_2e$ collide with
each other, they emit equivalent (virtual) photons with the
4--momenta $q_1$, $q_2$, energies $\omega_1$, $\omega_2$ and the
virtualities $Q_1^2=-q_1^2$, $Q_2^2=-q_2^2$.  Upon fusion, these
photons produce a {$\mu^{+}\mu^{-}$} pair with the total
four--momentum $q_1+q_2$ and the invariant mass squared $W^2 = (q_1
+q_2)^2$, besides we denote $(P_1+P_2)^2=4E^2=4M^2\,\gamma^2$,
$L=\ln{(\gamma^2)}$, $\alpha \approx 1/137$ and use the system of
units in which $c=1$ and $\hbar =1$.

The Born cross section of the process (\ref{2}) can be calculated
with a good accuracy using the equivalent photon approximation (EPA)
--- see, for example, Ref.~\cite{BGMS}. Let the numbers of
equivalent photons be $dn_1$ and $dn_2$. The most important
contribution to the production cross section stems from photons with
very small virtualities $Q_i^2 \ll {\mu}^2$ where $\mu$ is the muon
mass. In this very region the Born differential cross section
$d\sigma_{\rm B}$ for the considered process is related to the cross
section $\sigma_{\gamma\gamma}$ for the real $\gamma\gamma \to
\mu^{+}\mu^{-}$ process by the equation
\begin{equation}
d\sigma_{\rm B} = dn_1 dn_2 \, d\sigma_{\gamma \gamma} (W^2)\,,
\;\;\; W^2 \approx 4\,\omega_1\,\omega_2\,.
 \label{3}
\end{equation}
The number of equivalent photons are (see Eq.~(D.4) in
Ref.~\cite{BGMS})
 \begin{equation}
dn_i(\omega_i,Q_i^2) = {Z_i^2 \alpha \over \pi}\,\left(1-{\omega_i
\over E_i} \right)\, {d\omega_i \over
 \omega_i}\, \left( 1- {Q^2_{i\,\min} \over Q_i^2 } \right)
\,F^2(Q^2_i) \, {dQ^2_i\over Q^2_i} \,,
 \label{4}
\end{equation}
where
 \be
Q_i^2\ge Q^2_{i\,\min} = \omega_i^2/ \gamma^2
 \label{5}
 \ee
and $F(Q^2)$ is the nucleus electromagnetic form factor. It is
important that the integral over $Q^2$ converges fast at $Q^2
> 1/R^2$, were $R=1.2\,A^{1/3}\;\; {\rm fm}$ is the nucleus
radius and  $A\approx M/m_p$  ($R=7$ fm, $1/R= 28$ MeV for Au and
Pb). Since $Q^2_{\min} \lesssim 1/R^2$, the main contribution to the
cross section is given by virtual photons with energies $\omega_i
\lesssim \gamma/R$.

In calculation below we use the form factor in a simple approximate
form
\begin{equation}
 \label{26}
F(Q^2) = \frac{1}{1+Q^2/\Lambda^2}\,,
\end{equation}
which leads to
\begin{equation}
 \label{27}
dn_i(\omega_i) = {Z_i^2 \alpha \over \pi} \, f\left({\omega_i \over
\Lambda \gamma}\right)\, {d\omega_i\over \omega_i}
\end{equation}
with
\begin{equation}
 \label{28}
f(x)= (1 + 2\,x^2)\, \ln{\left({1\over x^2} +1\right)}\, -\, 2\,.
\end{equation}

Finally we obtain the Born cross section as a simple two dimension
integral:
 \be
\sigma_{\rm B} ={Z_1^2 Z_2^2 \alpha^2\over \pi^2 }\,
\int_{4\mu^2}^\infty\,{dW^2 \over W^2}\, G(W^2)\,\sigma_{\gamma
\gamma} (W^2)= {\left(Z_1 \alpha Z_2\alpha\right)^2 \over \pi
\mu^2}\, J(\gamma \Lambda/\mu)\,,
 \label{31}
\end{equation}
where
 \be
G(W^2)= \int\limits^{\omega_{\max}}_{\omega_{\min}} {d\omega \over
\omega}\; f\left({\omega \over \Lambda \gamma}\right)\,
f\left({W^2 \over 4\Lambda \gamma\omega}\right)\,.
 \label{32}
  \ee
It is easy to show that an accuracy of this calculation is
determined by the omitted terms of the order of $ \eta_1 =
\Lambda^2/(W^2 \, L)$, i.e. $\eta_1\sim 5 \% $  for the collisions
considered.

\begin{figure}[h]
\begin{center}
\includegraphics[width=5cm,angle=0]{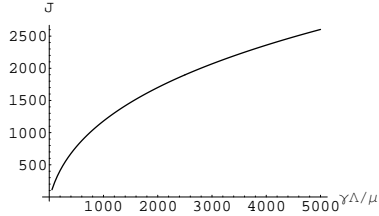}
 \caption{The function $J(\gamma\Lambda/\mu)$ from eq. (\ref{31})}
  \label{F:4}
\end{center}
\end{figure}

A numerical evaluation of the integrals in Eqs.~(\ref{31}),
(\ref{32}) yields the function $J(\gamma\Lambda/\mu)$ presented in
Fig. \ref{F:4}.

Let us now consider the probability of muon pair production $P_{\rm
B}(\rho)$ in collision of two nuclei at a fixed impact parameter
$\rho$. The Born cross section $\sigma_{\rm B}$ can be obtained by
the integration of $P_{\rm B}(\rho)$ over the impact parameters:
 \be
  \label{19}
 \sigma_{\rm B}=\int \,P_{\rm B}(\rho)\,d^2{\rho}\,.
 \ee
We calculate this probability in the LLA:
 \begin{equation}
P_{\rm B}(\rho) = \int dn_1 dn_2 \,\delta(\bm \rho_1 -\bm \rho_2-
\bm \rho)\,\sigma_{\gamma\gamma}(W^2)=
\frac{28}{9\pi^2}\,\frac{\left(Z_1 \alpha
Z_2\alpha\right)^2}{(\mu\rho)^2}\, \Phi(\rho)\,.
 \label{21}
\end{equation}
There are two scales in dependence of function $\Phi(\rho)$ on
$\rho$:
 \bea
\Phi(\rho)=& \left(4\ln{ \fr{\gamma}{\mu\rho}}
+\ln{\fr{\rho}{R}}\right)\, \ln{\fr{\rho}{R}}\; \;\;\;\mbox{at}\;\;&
R \ll \rho \le \gamma/\mu\,,
 \label{22}\\
\Phi(\rho)=&  \left(\ln{\fr{\gamma^2}{\mu^2  \rho R}}\right)^2
\;\;\;\;\;\;\;\;\;\;\;\;\;\;\;\;\;\;\;\mbox{at}\;\;&\gamma/\mu \le
\rho \ll \gamma^2/(\mu^2 R)\,.
  \label{23}
 \eea
We compare Eqs. for $\Phi(\rho)$ with the numerical calculations
based on the exact matrix element calculated with the approach
outlined in\cite{HTB1994}. There is a good agreement for the Pb-Pb
collisions: the discrepancy is less then 10 \% at $\mu \rho > 10$
and it is less then 15 \%  at $\mu \rho > 2\mu R =7.55$.

\section{Coulomb and unitarity corrections}

The Coulomb correction corresponds to Feynman diagram of Fig.
\ref{F:2}. Due to restriction of transverse momenta of additional
exchange photons on the level of $1/R$, the effective parameter of
the perturbation series is not $(Z\alpha)^2$ but
$(Z\alpha)^2/(R\mu)^2$. Besides, the contribution of the additional
photons is suppressed by logarithmic factor. Indeed, the cross
section for two--photon production mechanism is proportional to
$L^3$, while the cross section for multiple-photon production
mechanism is proportional to $L^2$. Therefore, the real suppression
parameter is of the order of $\eta_2= (Z\alpha)^2/[(R \mu )^2 L]$,
which corresponds to the Coulomb correction less then $1\,$\%.

The unitarity correction $\sigma_{\rm unit}$ to one muon pair
production corresponds to the exchange of light-by-light blocks
between nuclei (Fig. \ref{F:3}). We start with more general process
--- the production of one $\mu^+\mu^-$ pair and $n$
electron-positron pairs ($n\geq 0$) in collision of two
ultra-relativistic nuclei
 \be
Z_1+Z_2 \to Z_1+Z_2+\mu^+\mu^- +n\, (e^+e^-)
 \label{37}
 \ee
taking into account the unitarity corrections which corresponds to
the exchange of the blocks of light-by-light scattering via the
virtual lepton loops. The corresponding cross section
$d\sigma_{1+n}$ can be calculated by a simple generalization of the
results obtained in paper~\cite{BGKN} for the process without muon
pair production: $Z_1+Z_2 \to Z_1+Z_2+n\, (e^+e^-)$. Our result is
the following
  \be
{d\sigma_{1+n}\over d^2\rho} =P_{\rm B} (\rho) \,
{[\bar{n}_e(\rho)]^n\over n!}\,{\rm e}^{-\bar{n}_e(\rho)}\,,
 \label{41}
 \ee
where $\bar{n}_e(\rho)$ is the average number of the $e^+e^-$ pairs
produced in collisions of two nuclei at a given impact parameter
$\rho$.

In particular, the cross section for the one $\mu^+\mu^-$ pair
production including the unitarity correction is
 \be
\sigma_{1+0} = \int_{2R}^\infty P_{\rm B} (\rho)\,{\rm
e}^{-\bar{n}_e(\rho)} \,d^2\rho\,.
 \label{42}
 \ee
This expression can be rewritten in the form $\sigma_{1+0} =
\sigma_{\rm B} +\sigma_{\rm unit}\,,$ where
 \be
\sigma_{\rm B}=\int_{2R}^\infty P_{\rm B} (\rho)\,\,d^2\rho
 \label{44}
 \ee
is the Born cross section discussed in Sect. 2 and
 \be
\sigma_{\rm unit}=- \int_{2R}^\infty \left[1-{\rm
e}^{-\bar{n}_e(\rho)}\right]\, P_{\rm B} (\rho)\,d^2\rho
 \label{45}
 \ee
corresponds to the unitarity correction for one muon pair
production.

In LLA we find $\sigma_{\rm unit}\sim -1.2$ barn for the Pb-Pb
collisions at LHC, which corresponds approximately to $-50$\% of the
Born cross section. It is seen that unitarity corrections are large,
in other words, the exclusive production of one muon pair differs
considerable from its Born value

\section{Inclusive production of one $\mu^+\mu^-$ pairs}

The experimental study of the exclusive muon pair production seems
to be a very difficult task, because this process requires that the
muon pair should be registered without any electron-positron pair
production including $e^\pm$ emitted at very small angles.
Otherwise, the corresponding cross section will be close to the Born
cross section.

Indeed, it is clearly seen from the expression for $\sigma_{1+n}$
that after summing up over all possible electron pairs we obtain the
Born cross section $\sum_{n=0}^{\infty}\, \sigma_{1+n} = \sigma_{\rm
B}$. Therefore, there is a very definite prediction: the inclusive
muon pair production coincides with the Born limit. This direct
consequence of calculations taking into account strong field effects
may be more easier for an experimental test that the prediction for
cross sections of several $e^+e^-$ pair production.

 \section{Conclusion}

The exclusive production of one $\mu^+\mu^-$ pair in collisions of
two ultra-relativistic nuclei is considered. We present the simple
method for calculation of the Born cross section for this process.

Then we found that the Coulomb corrections to the this cross section
are small while the unitarity corrections are large. This is in
sharp contrast to the exclusive $e^+e^-$ pair production where the
Coulomb corrections to the Born cross section are large while the
unitarity corrections are small.

We calculated also the cross section for the production of one
$\mu^+\mu^-$ pair and several $e^+e^-$ pairs in LLA. Using this
cross section we found that the inclusive production of $\mu^+\mu^-$
pair coincides in this approximation with its Born value.



\begin{thebibliography}{99}


\bibitem{BHTSK}
G.~Baur, K.~Hencken, D.~Trautmann, S.~Sadovsky, Yu.~Kharlov, {\it
Phys. Rep.} {\bf 364}, 359 (2002).

\bibitem{Landau}
L.D.~Landau, E.M.~Lifshitz, {\it Phys. Zs. Sowjet} {\bf 6}, 244
(1934); G.~Racah, {\it Nuovo Cimento} {\bf 14}, 93 (1937).

\bibitem{ISS-99}
D.Yu.~Ivanov, A.~Schiller, V.G.~Serbo, {\it Phys. Lett.} {\bf B
454}, 155 (1999).

\bibitem{LMS-02}
R.N.~Lee, A.I.~Milstein, V.G.~Serbo, {\it Phys. Rev.} {\bf A 65},
022102-1 (2002).

\bibitem{BGKN}
E.~Barto$\check{\rm s}$, S.R.~Gevorkyan, E.A.~Kuraev, N.N.~Nikolaev,
{\it Phys. Lett.} {\bf B 538}, 45 (2002) and hep-ph/0204327.

\bibitem{BGMS}
V.~M.~Budnev, I.~F.~Ginzburg, G.~V.~Meledin, V.~G.~Serbo, {\it Phys.
Rep.} {\bf C 15}, 181 (1975).

\bibitem{HTB1994}
K.~Hencken, D.~Trautmann, G.~Baur, {\it Phys. Rev.} {\bf A51},1874
(1995).

\end{thebibliography}
\end{document}